# Performance of Resistive Plate Chambers installed during the first long shutdown of the CMS experiment

**M. Shopova[a*], A. Aleksandrov[a], R. Hadjiiska[a], P. Iaydjiev[a], G. Sultanov[a], M. Rodozov[a], S. Stoykova[a], Y. Assran[b], A. Sayed[b], A. Radi[b], S. Aly[b], G. Singh[c], M. Abbrescia[d], G. Iaselli[d], M. Maggi[d], G. Pugliese[d], P. Verwilligen[d], W. Van Doninck[e], S. Colafranceschi[f], A. Sharma[f], L. Benussi[g], S. Bianco[g], D. Piccolo[g], F. Primavera[g], A. Cimmino[h], S. Crucy[h], A. A. O. Rios[h], M. Tytgat[h], N. Zaganidis[h], M. Gul[h], A. Fagot[h], V. Bhatnagar[i], J. Singh[i], R. Kumari[i], A. Mehta[i], A. Ahmad[j], I. M. Awan[j], H. Shahzad[j], H. Hoorani[j], M. I. Asghar[j], S. Muhammad[j], W. Ahmed[j], M. A. Shah[j], S. W. Cho[k], S. Y. Choi[k], B. Hong[k], M. H. Kang[k], K. S. Lee[k], J. H. Lim[k], S. K. Park[k], M. S. Kim[l], I. B. Laktineh[m], F. Lagarde[m], M. Gouzevitch[m], G. Grenier[m], I. Pedraza[n], S. Carpinteyro Bernardino[n], C. Uribe Estrada[n], S. Carrillo Moreno[o], F. Vazquez Valencia[o], L. M. Pant[p], S. Buontempo[q], N. Cavallo[q], F. Fabozzi[q], I. Orso[q], L. Lista[q], S. Meola[q], M. Merola[q], P. Paolucci[q], F. Thyssen[q], G. Lanza[q], M. Esposito[q], A. Braghieri[r], A. Magnani[r], C. Riccardi[r], P. Salvini[r], I. Vai[r], P. Vitulo[r], P. Montagna[r], Y. Ban[s], S. J. Qian[s], M. Choi[t], Y. Choi[u], J. Goh[u], D. Kim[u], A. Dimitrov[v], L. Litov[v], P. Petkov[v], B. Pavlov[v], I. Bagaturia[w], D. Lomidze[w], C. Avila[x], A. Cabrera[x], J. C. Sanabria[x], I. Crotty[y], and J. Vaitkus[z]**

[a] *Bulgarian Academy of Sciences, Inst. for Nucl. Res. and Nucl. Energy, Tzarigradsko shaussee Boulevard 72, BG-1784 Sofia, Bulgaria.*
[b] *Egyptian Network for High Energy Physics, Academy of Scientific Research and Technology, 101 Kasr El-Einy St. Cairo, Egypt.*
[c] *Chulalongkorn University, Department of Physics, Faculty of ScienceName of Institute, Payathai Road, Phatumwan, Bangkok, THAILAND – 10330.*
[d] *INFN, Sezione di Bari, Via Orabona 4, IT-70126 Bari, Italy.*
[e] *Vrije Universiteit Brussel, Boulevard de la Plaine 2, 1050 Ixelles, Belgium.*
[f] *Physics Department CERN, CH-1211 Geneva 23, Switzerland.*
[g] *INFN, Laboratori Nazionali di Frascati (LNF), Via Enrico Fermi 40, IT-00044 Frascati, Italy.*
[h] *Ghent University, Department of Physics and Astronomy, Proeftuinstraat 86, B-9000 Gent, Belgium.*
[i] *Department of Physics, Panjab University, Chandigarh Mandir 160 014, India.*
[j] *National Centre for Physics, Quaid-i-Azam University, Islamabad, Pakistan.*
[k] *Korea University, Department of Physics, 145 Anam-ro, Seongbuk-gu, Seoul 02841, Republic of Korea.*

[*] Corresponding author.

[l] *Kyungpook National University,*
*80 Daehak-ro, Buk-gu, Daegu 41566, Republic of Korea.*
[m] *Universite de Lyon, Universite Claude Bernard Lyon 1, CNRS-IN2P3, Institut de Physique Nucleaire de Lyon,*
*Villeurbanne, France.*
[n] *Benemerita Universidad Autonoma de Puebla,*
*Puebla, Mexico.*
[o] *Universidad Iberoamericana,*
*Mexico City, Mexico.*
[p] *Nuclear Physics Division Bhabha Atomic Research Centre,*
*Mumbai 400 085, INDIA.*
[q] *INFN, Sezione di Napoli, Complesso Univ. Monte S. Angelo,*
*Via Cintia, IT-80126 Napoli, Italy.*
[r] *INFN, Sezione di Pavia,*
*Via Bassi 6, IT-Pavia, Italy.*
[s] *School of Physics, Peking University,*
*Beijing 100871, China.*
[t] *University of Seoul,*
*163 Seoulsiripdae-ro, Dongdaemun-gu, Seoul, Republic of Korea.*
[u] *Sungkyunkwan University,*
*2066 Seobu-ro, Jangan-gu, Suwon-si, Gyeonggi-do, Republic of Korea.*
[v] *Faculty of Physics, University of Sofia,*
*5, James Bourchier Boulevard, BG-1164 Sofia, Bulgaria.*
[w] *Tbilisi University,*
*1 Ilia Chavchavadze Ave, Tbilisi 0179, Georgia.*
[x] *Universidad de Los Andes,*
*Apartado Aereo 4976, Carrera 1E, no. 18A 10, CO-Bogota, Colombia.*
[y] *Dept. of Physics, Wisconsin University,*
*Madison, WI 53706, United States.*
[z] *Vilnius University,*
*Vilnius, Lithuania.*

*E-mail*: mariana.vutova@cern.ch

ABSTRACT: The CMS experiment, located at the CERN Large Hadron Collider, has a redundant muon system composed by three different detector technologies: Cathode Strip Chambers (in the forward regions), Drift Tubes (in the central region) and Resistive Plate Chambers (both its central and forward regions). All three are used for muon reconstruction and triggering. During the first long shutdown (LS1) of the LHC (2013-2014) the CMS muon system has been upgraded with 144 newly installed RPCs on the forth forward stations. The new chambers ensure and enhance the muon trigger efficiency in the high luminosity conditions of the LHC Run2. The chambers have been successfully installed and commissioned. The system has been run successfully and experimental data has been collected and analyzed. The performance results of the newly installed RPCs will be presented.

## Contents

## 1. Instroduction

The Compact Muon Solenoid (CMS) [1] is a multipurpose detector operating at the Large Hadron Collider (LHC) at CERN, which has been successfully collecting data since the start of the first physics run period in 2009. Foreseeing the requirements of the high luminosity LHC physics run, several major upgrades of the CMS experiment were completed. One of the major upgrades was done in the CMS muon system, where three different types of gaseous detectors are used to identify and characterize muons - Drift Tubes (central region), Cathode Strip Chambers (forward regions) and Resistive Plate Chambers (both in central and forward regions). During the first long shutdown (LS1) of the LHC (2013 - 2014), the CMS muon upgrade collaboration added 144 new double-gap RPC detectors, thus completing the 4th forward stations [2]. Adding these stations, referred to as RE4 stations, increased the overall robustness of the CMS muon spectrometer and improved the trigger efficiency in the End-Cap region with pseudorapidity in the range $1.2 < |\eta| < 1.6$, shown in Fig. 1.

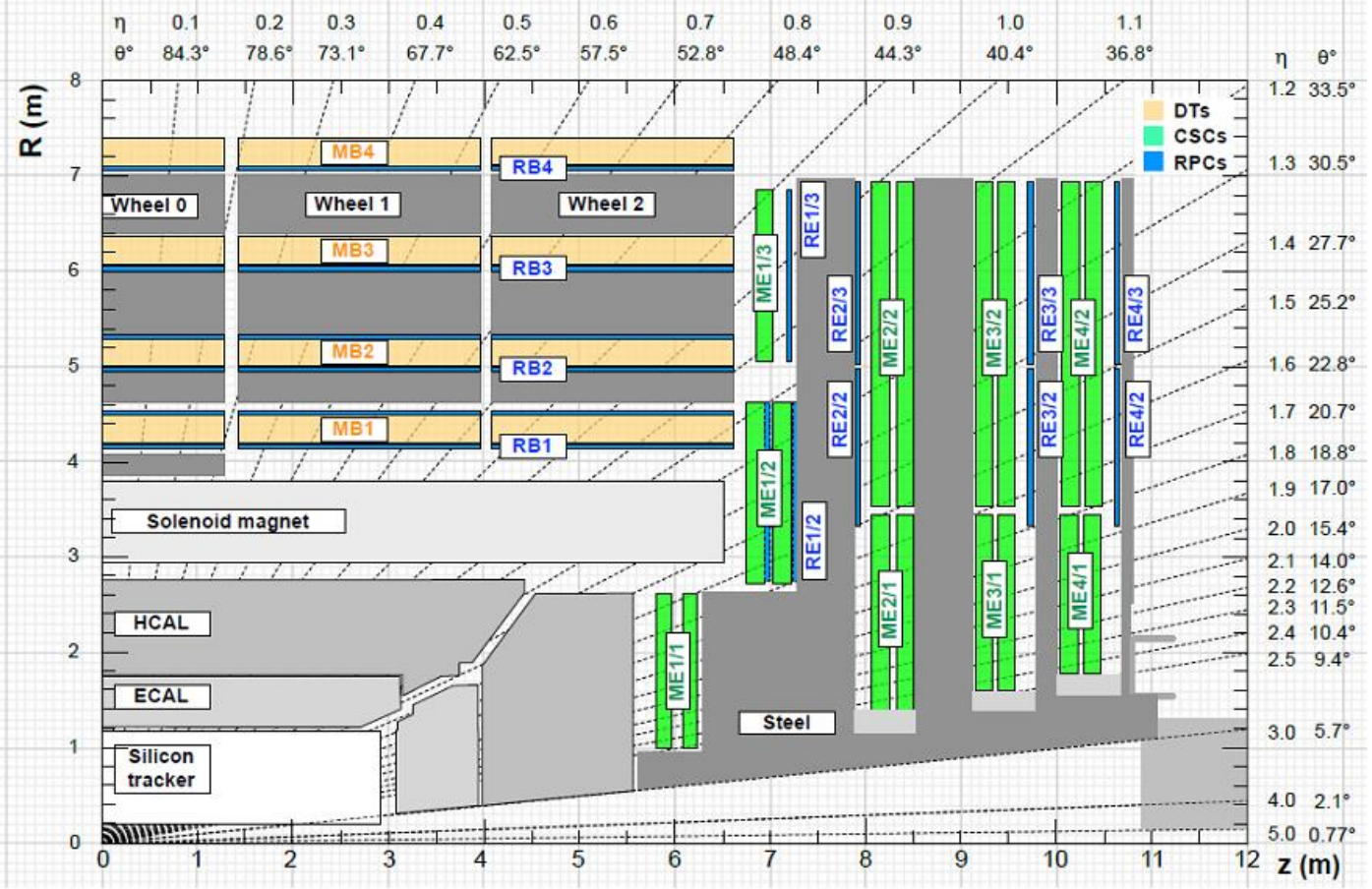

**Figure 1.** Longitudinal layout of one quadrant of the CMS detector which shows the enhancement in trigger efficiency with 4th End-Cap (RE4) in the pseudorapidity region $1.2 < |\eta| < 1.6$.

## 2. Chamber design and performance

The newly installed RE4 chambers inherit their design from the already existing RPC Endcap chambers, shown in Fig. 2 [3]. The detector relies on 2 mm trapezoidal shaped High Pressure Laminate (HPL) gas gaps, organized in a double-layer configuration with a copper strip readout panel placed in between. The HPL sheets resistivity is of the order of 1 – 6*$10^{10}$ Ωcm. A chamber is made of three kinds of different gap geometries – one large gas gap is used at the bottom layer, while the top layer is segmented into two parts. This configuration is used to simplify the signal cable routing.

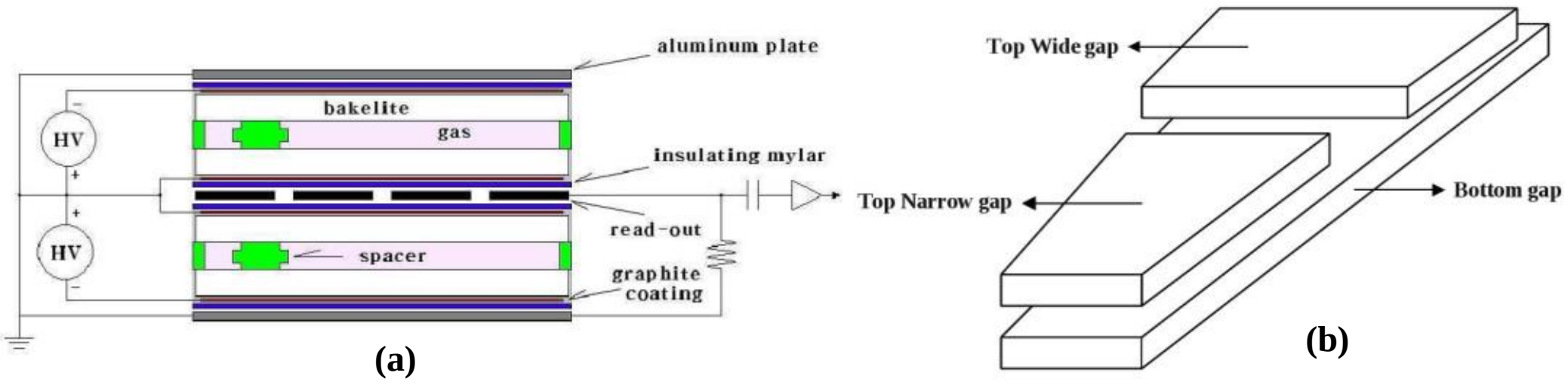

**Figure 2.** Schematic representation of the double-layer layout of the endcap chambers (a) and configuration of the gas gaps of the endcap chambers (b).

The readout strips are segmented in three η-partitions with increasing strip pitch from 1.5 cm to 4 cm [3].Since each partition has 32 strips, there are a total of 96 strips per chamber. Coaxial cables are used to connect strips to Adapter Boards which are linked to 3 Front-End Boards (FEBs) per chamber. There is 1 Distribution Board (DB) for the electronics control for each chamber. Link Boards (LBs) are the main components of the off-detector electronics. They receive from the FEBs signals in LVDS (Low-Voltage Differential Signaling) standard. Synchronization with the LHC clock and transmission of the signals to Trigger Electronics in the control room is performed by the LBs. There is a Control Board (CB) in each LB crate that drives the crate, provides inter-crate communication and takes care of the connection to the readout and trigger systems.

### 2.1 Evaluation of the optimal High Voltage working point

The efficiency of the RPC detectors strongly depends on the applied high voltage. Due to temperature and pressure variation, the applied high voltage is corrected and gives the effective voltage. The correction is performed according to the formula below [4]:

$$HV_{eff} = HV_{app} \cdot \frac{p_0}{p} \cdot \frac{T}{T_0} \quad (1),$$

Where the $T_0$ (293 K) and $P_0$ (1010 mbar) are the reference temperature and pressure.

For each chamber a High Voltage (HV) scan is performed in order to determine the optimal HV working point. The HV scan allows to measure the Efficiency as a function of the Effective voltage. The results of the scan are fitted to a sigmoid function and the HV point where the efficiency curve reaches a plateau is evaluated, as shown in Fig. 3a. The HV50 used in the fitting sigmoid function [4] is defined as the high voltage at which every roll reaches 50% of the plateau efficiency. The working point (WP) is the voltage applied to nominal operation. It is defined as WP = Knee + 120V for every RPC chamber. Here Knee is the high voltage point on the sigmoid where the efficiency is 95% of the plateau efficiency.

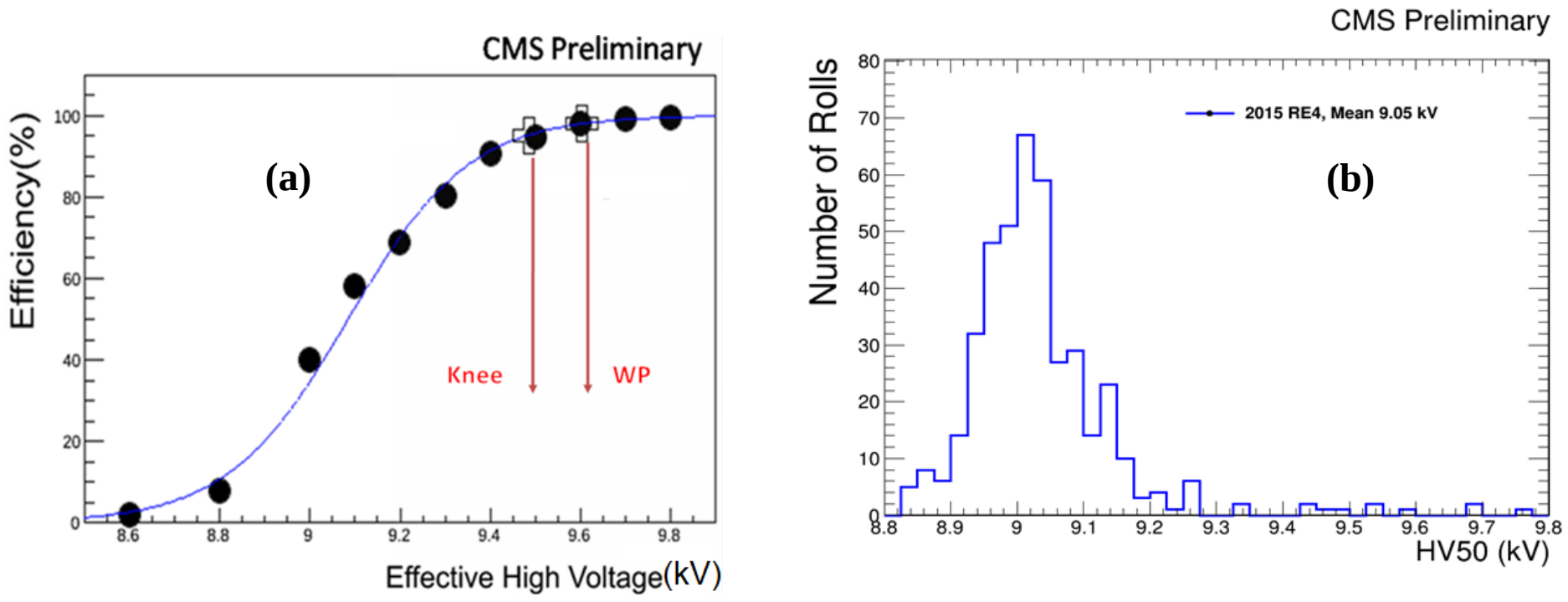

**Figure 3.** A typical HV scan showing the working point and the knee for one of the RE4 chambers (a) and the high voltage distribution at 50% efficiency for the $4^{th}$ station of the endcaps (b).

The HV scan was done with 2015 data. The width and the peak of the HV50 distribution shown in Fig. 3b depend mostly on the operational conditions and construction specifications such as spacer sizes. The spacers are the supports that create the RPC gaps in the chambers.

### 2.2 Efficiency and hit rates

The RPCs are installed on two rings per station and every chamber is subdivided in 3 eta partitions called rolls. The detector performance is monitored via the occupancy distribution of the chambers. The cross-sectional view of both positive and negative RE4 stations is shown in Fig. 4a.

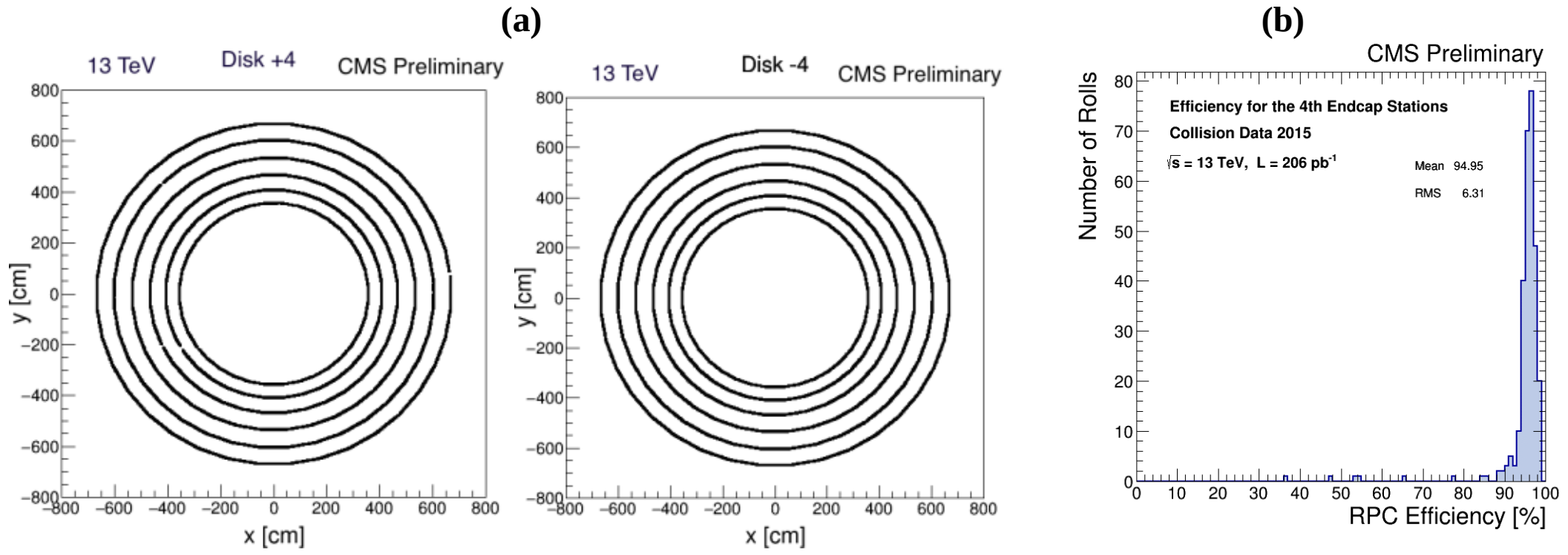

**Figure 4.** Reconstructed muon hits (XY view occupancy) on the forth positive and negative endcap stations (a) and overall efficiency distribution of the fourth endcap stations at 3.8 Tesla (b).

The black points show the position of the reconstructed hits in the middle of the signal electrodes (strips). It is evident from plots that there are no inactive channels in the newly installed chambers.

The RPC efficiency is calculated using the segment extrapolation method, explained in detail in [5]. The overall efficiency distribution based on the analysis with proton-proton collision data at sqrt(s) = 13 is shown in Fig.4b. It is obtained also after the new HV working point have been deployed. After setting the new WPs, improvement of about 0.7% is observed. Few chambers with low efficiency in the distribution correspond to known hardware problems. The RE4 efficiency mean value is 94.95% which is in good agreement with expectations. The history of the overall RPC efficiency over the period of 2015 data taking is shown in Fig. 6a.

The efficiency and the measured hit rate for each detector unit are shown in 2D maps in Fig. 5. The X-axis corresponds to the chamber number – there are 36 chambers per ring, while the Y-axis corresponds to the ring number and the names of the detector units. The analysis is based on data taken during the 2015 proton-proton collisions.

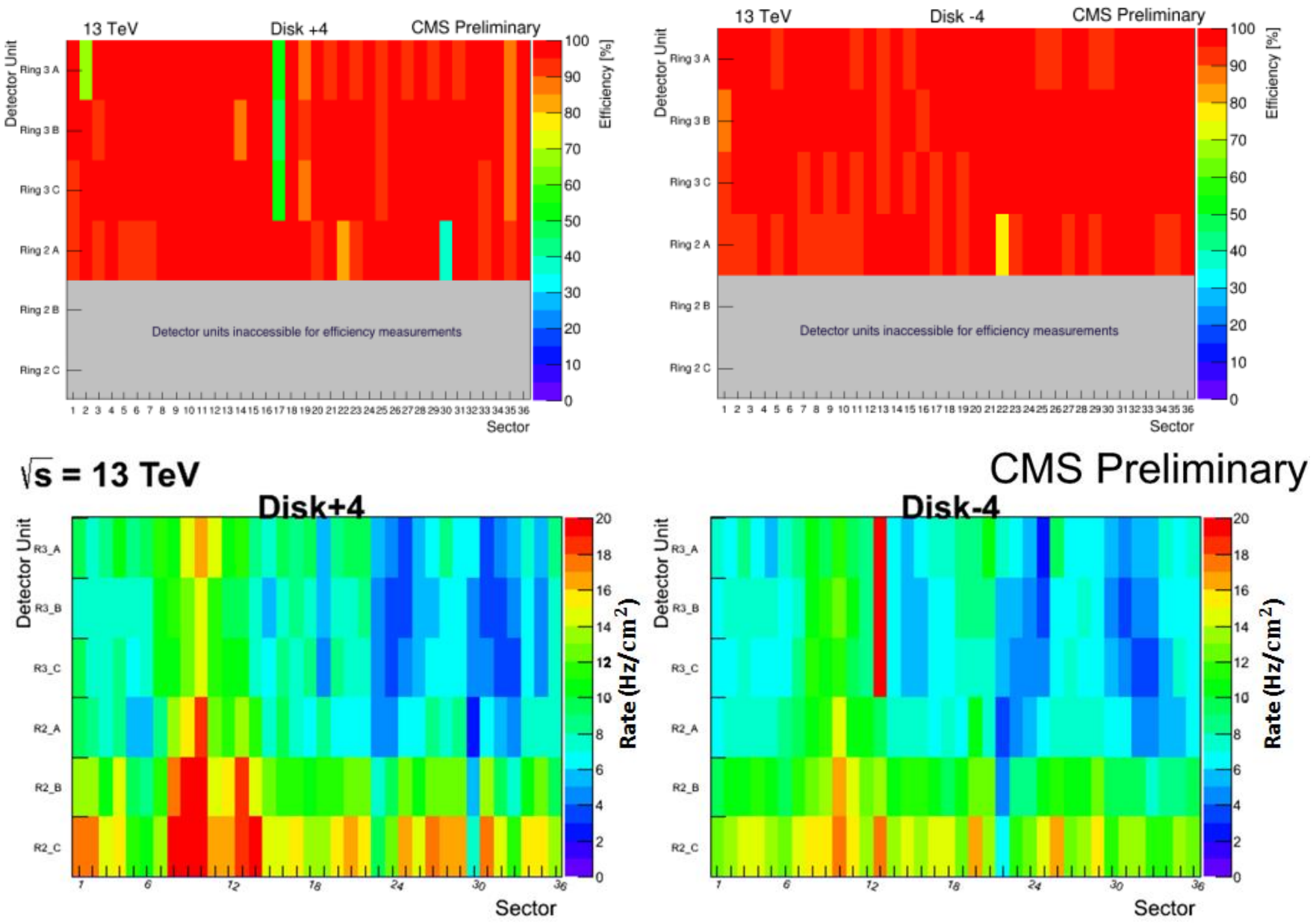

(a)

(b)

**Figure 5.** 2D Efficiency map (a) and 2D Hit Rate map (b) of the RPCs installed on the positive and negative fourth endcap stations.

In Fig. 5a the grey entries correspond to the detector units excluded from efficiency calculation because the software algorithm is not effective for them due to geometrical constrains. The blue and the green colors correspond to the lower efficiency values measured for detector units which are partially masked.

The RPC hit rate (in Hz/cm$^2$) is measured for a run at average instantaneous luminosity of $4.5*10^{33}$ cm$^{-2}$s$^{-1}$. In Fig 5b the blue and the violet colors correspond to the lower rates, while the yellow, the orange and the red colors correspond to high background level. The average hit rate for the shown maps is ~10 Hz/cm$^2$. It is in agreement with previous measurements and expectation from MC simulation, that the higher rate is observed for higher eta regions [5].

### 2.3 Cluster size

A consecutive set of strips, each collecting an induced charge defines a cluster. The number of the strips in the cluster gives the Cluster size. Fluctuations in the efficiency and cluster size history distributions shown in Fig. 6 are due to the performed scans (HV scan in the middle of June & threshold scan in beginning of October).

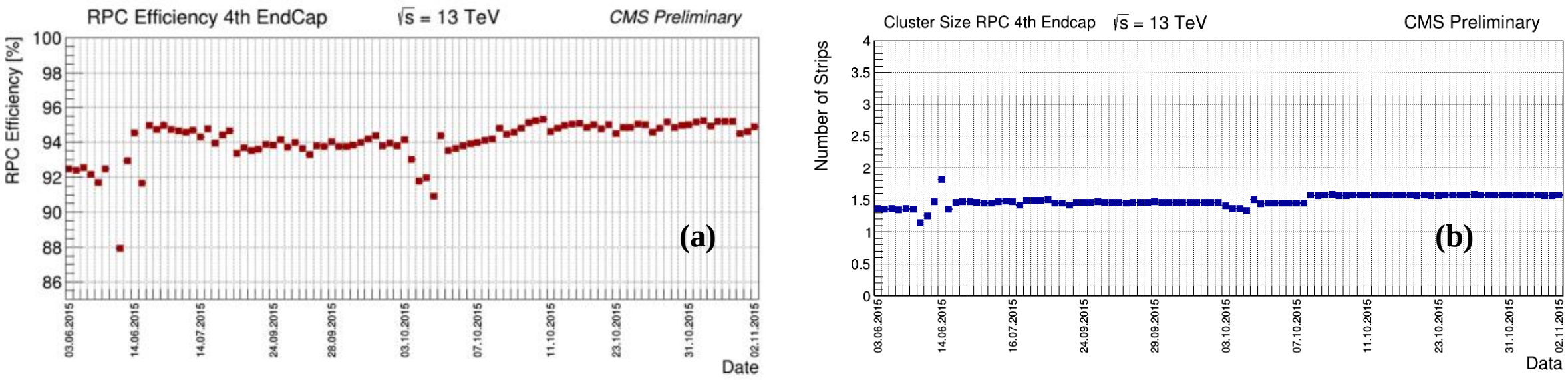

**Figure 6.** History of the overall RPC Efficiency (a) and the mean Cluster size (b) of the newly installed RPCs on the positive and negative endcap station for the 2015 physics data taking.

The system has higher efficiency and its performance becomes more stable after deploying the new HV working point (October 2015). The dependence of the RPC efficiency on the atmospheric pressure in the cavern is compensated with automatic corrections of the applied HV. The average cluster size is persistently below 2, which is in good agreement with the expectations [1] and the performance of all the other RPCs from the CMS muon system.

## 3. Conclusions

During the first long shutdown (LS1) of LHC the RPC collaboration built, installed and commissioned 144 new RPC detectors in the CMS experiment. The system of newly installed chambers has run successfully with average efficiency of 94.95%, average cluster size persistently below 2 and average rate of ~10 Hz/cm$^2$ in agreement with expectations and simulations. The quality of the experimental data taken during the first year of physics data taking is high and in agreement with expectations.

## Acknowledgments

The authors would like to thank everyone in the CERN accelerator departments for the excellent performance of the LHC machine. We would also like to thank our colleagues from the CMS muon system and the RPC collaboration.